Interface, bulk and surface structure of heteroepitaxial α-MnTe films grown on GaAs(111).

Sara Bey[1], Maksym Zhukovskyi[2], Tatyana Orlova[2], Shelby Fields[3], Valeria Lauter[4], Haile Ambaye[4], Anton Ievlev[3], Steven P. Bennett[4], Xinyu Liu[1], Badih A. Assaf[1]

[1] Department of Physics and Astronomy, University of Notre Dame, Notre Dame IN, 46556, USA.
[2] Notre Dame Integrated Imaging Facility, University of Notre Dame, Notre Dame, IN, 46556, USA
[3] Materials Science and Technology Division, U.S. Naval Research Laboratory, Washington DC, 20375, USA
[4] Neutron Sciences Directorate, Oak Ridge National Laboratory, Oak Ridge, TN, 37831, USA
[5] Center for Nanophase Materials Science, Oak Ridge National Laboratory, Oak Ridge, TN, 37831, USA

**Abstract.** Epitaxial MnTe films have recently seen a spur of research into their altermagnetic semiconducting properties. However, those properties may be extremely sensitive to structural and chemical modifications. We report a detailed investigation of the synthesis of the altermagnet α-MnTe on GaAs(111) which reveals the bulk defect structure of this material, the mechanism by which it releases strain from the underlying substrate and the impact of oxidation on its surface. X-ray diffraction measurements show that α-MnTe layers with thicknesses spanning 45nm to 640nm acquire lattice parameters different from bulk mostly due to thermal strain caused by the substrate rather than strain from the lattice mismatch. Through high resolution transmission electron microscopy (TEM) measurement, we then unveil a misfit dislocation array at the interface, revealing the mechanism by which lattice strain is relaxed. TEM also reveals a stacking fault in the bulk, occurring along a glide plane parallel to the interface. The combination of TEM with polarized neutron reflectometry measurements finally reveals the impact of oxidation on the chemistry of the surface of uncapped MnTe. Our findings highlight the subtle role of epitaxy in altering the structure of α-MnTe providing potential opportunities to tune the altermagnetic properties of this material.

**Introduction**

The process of synthesizing thin films of a given material on poorly matched substrate introduces stresses that can impact the pristine structure of the material causing defects and dislocations to form. Various recent works, report the electrical, magnetic and optical properties of a new class of magnetic materials referred to as altermagnets and have relied on epitaxial thin films whose structural properties are not yet understood. Altermagnets are compensated magnetic materials, where low magnetocrystalline symmetry allows spin splitting as well as various other properties common to ferromagnets to be observed [1–4]. For instance, in the leading altermagnetic material α-MnTe, the anomalous Hall effect has been reported in single crystal and epitaxial thin films [5–9], despite the material having a nearly vanishing out-of-plane magnetization. And in CrSb, a magnetic splitting of the electronic bands has been seen in photoemission, also without a net magnetization. [10–12]. To date, a number of altermagnetic materials have already been proposed, predicted or identified [2,13–16].

α-MnTe has been grown by molecular beam epitaxy on lattice matched InP(111) and SrF$_2$(111) as well as on lattice mismatched GaAs(111) and Al$_2$O$_3$(0001) [5,8,9,17,18]. SrF$_2$(111) is by far the ideal substrate to grow this material, due to its lattice match and its matching coefficient of thermal expansion (CTE) [9,19,20]. InP(111) and GaAs(111) both have poorly matched CTEs [21]. There are however unexpected findings of a tunable anomalous Hall effect in α-MnTe grown on GaAs(111) that make this heterointerface unique and of interest to investigate more thoroughly [9].

Here, we report a detailed study of the crystallographic properties and the defect structure of α-MnTe/GaAs(111) with focus on results from high-resolution X-ray diffraction (HR-XRD), polarized neutron reflectometry and transmission electron microscopy (TEM). We combine HR-XRD, with cross-sectional TEM to probe the structure of the surface, bulk, and interface of α-MnTe films grown on GaAs(111). We find that α-MnTe grows thermally-strained on GaAs(111), up to a thickness of 640nm. We additionally reveal that despite the large lattice mismatch between the two materials, α-MnTe films do not exhibit any visible screw dislocations propagating through the bulk, but instead release strain through an interfacial array of misfits that is confined to the first few monolayers. We also identify a series of unconventional planar dislocations that arise in the bulk of the layer along glide planes parallel to the interface. Lastly, we reveal the impact of oxidation on the surface of this material, arguing the importance of ruling out its role when studying the electrical and magnetic properties of this material. Our work provides an essential understanding of the defect structure and chemistry of the MBE grown α-MnTe. It delivers crucial knowledge relevant to interpret experiments that probe the fundamental characteristics this prototypical altermagnet.

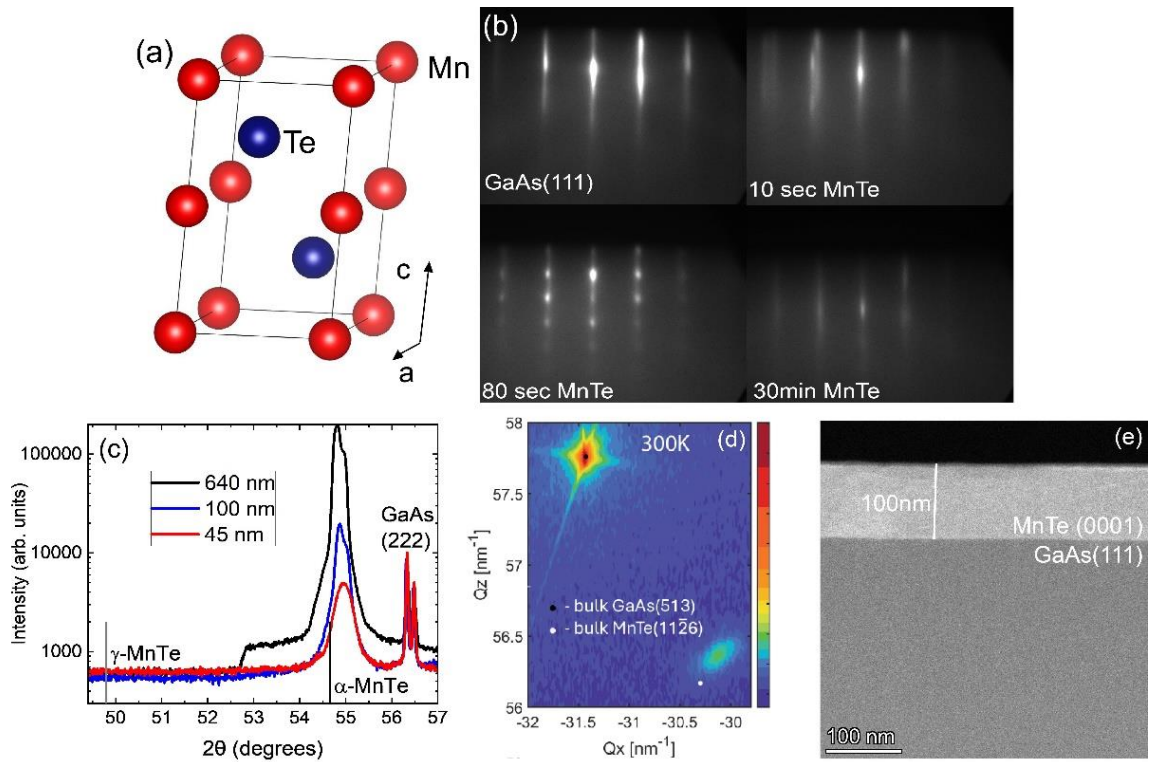

**FIG 1**. (a) Crystal structure of α-MnTe. (b) Reflection high energy electron diffraction (RHEED) images taken at different times during the growth. (c) Specular X-ray diffraction pattern about the (0004) peak of α-MnTe. Three samples of varying thickness are measured. (d) Reciprocal space map of the $(11\bar{2}6)$ peak of a 100nm α-MnTe grown on GaAs(111). The white dot represents the location of the expected Bragg peak for bulk α-MnTe (e) Low magnification scanning transmission electron microscopy image of the same film in (d) along a $[1\bar{1}0]$ projection.

**Results**

**Synthesis and X-ray diffraction**

α-MnTe is grown by molecular beam epitaxy on GaAs(111)B. Initially, the GaAs substrates are thermally annealed to desorb the native oxide at $T_s$=570°C, then annealed again under a Te flux at $T_s$=600°C, following a routine procedure used in previous work [22]. α-MnTe is then synthesized by co-evaporating elemental Mn and Te from K-cells at a relative beam equivalent pressure of 1 to 5, under Te rich conditions. The substrate temperature is maintained at 400°C - 450°C during the growth. It is calibrated by the desorption process of the GaAs surface native oxide. The substrate temperature and relative Mn:Te flux are chosen carefully to eliminate the formation of zincblende MnTe. The growth of MnTe is monitored by reflection high energy electron diffraction (RHEED) shown at different times in Fig. 1(b). After Te treatment, we recover the pristine RHEED pattern in GaAs(111). The very first MnTe layer grows with the different lateral lattice parameter as the GaAs substrate as seen in the RHEED image taken 10s after the Mn shutter is opened. Above ~2 ML, the streaks break up into spots, indicating the initiation of 3D growth as seen in the image taken at 80s into the growth. After deposition of 10-20nm MnTe, a (1x1) RHEED pattern is observed (30min into the growth), and is characterized by sharp fundamental streaks. The appearance of double streaks as soon as 10s (Fig. 1(b)) after the shutters open resembles what is reported in MBE grown transition metal dichalcogenides [23,24]. It suggests the formation of a transient van-der-Waals layer as the growth is initiated. That layer gets absorbed at later times, and does not show up in any ex-situ characterization measurements. The dynamics of the RHEED patterns indicate a quick strain relaxation mechanism that we investigate in detail later on.

Specular x-ray diffraction measurements near the (0004) Bragg peak of MnTe shown in Fig. 1(b) confirm the formation of the α-phase where altermagnetism is expected. Regardless of thickness, we observe a (0004) peak at an angle higher than what is expected from the bulk lattice parameters [25]. Even the 640nm film remains partially strained and does not yield the lattice parameter expected for bulk.

Off-specular reciprocal space maps (RMS) near the (513) GaAs Bragg peak reveal the origin of this strain. The RSM is shown for the 100nm sample at 300K in Fig. 1(c). We determine the lattice parameters of the 100nm MnTe layer - a=4.178Å and c=6.6901Å - using the $(11\bar{2}6)$ Bragg peak seen in Fig. 1(c). We find an in-plane lattice parameter a larger than in bulk ($a_{bulk}$=4.148Å) [25], despite GaAs having an in-plane lattice spacing ($a_{[110]}$=3.9975Å) that is smaller than bulk MnTe. We conclude that the dominant origin of this strain cannot be the lattice mismatch between the two materials, but rather the mismatch between their CTE. The results seen in X-ray diffraction raise the question of whether and how lattice strain is relieved.

We next characterize the substrate-layer interface by carrying out cross-sectional TEM measurements. All high-resolution cross-sectional scanning TEM (STEM) images are acquired using a double tilt holder and probe corrected Spectra 30-300 transmission electron microscope (Thermo Fisher Scientific, USA) equipped with a field emission gun, operated at 300 kV. The images are acquired using Panther STEM detector (Thermo Fisher Scientific, USA) in high-angle, annular dark field mode (HAADF) and bright field mode (BF). For compositional analysis, energy-dispersive X-ray spectroscopy (EDX) maps were obtained in STEM mode using the Super-X EDX system. TEM samples were prepared by focused ion beam etching using the standard lift-out technique. Fig. 1(e) shows a low magnification STEM image, highlighting the uniformity of the layer on the microscopic scale. No threading dislocations are seen penetrating through the bulk, despite the large lattice mismatch between GaAs and MnTe.

**Interface structure**

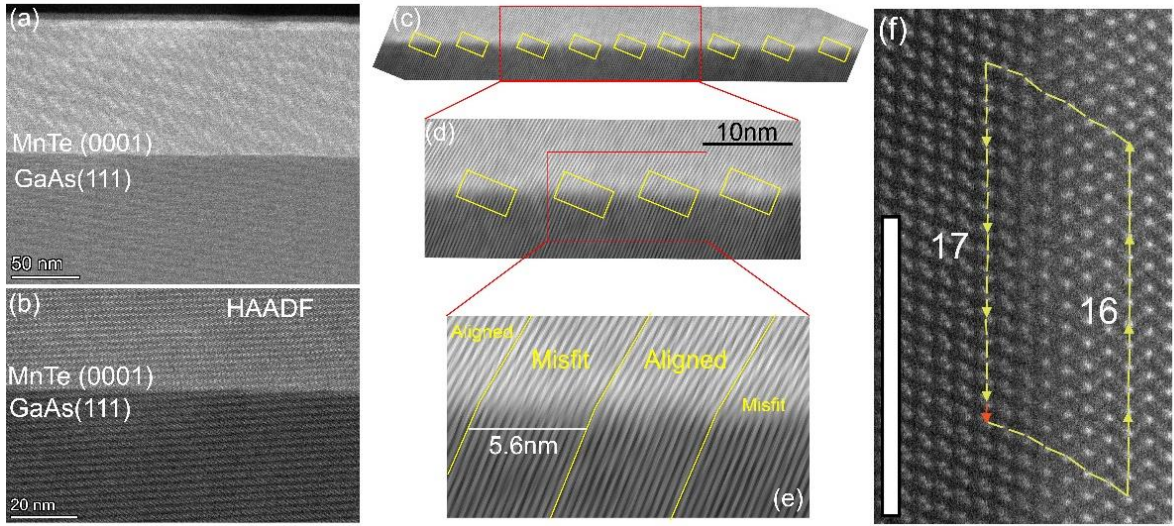

**FIG 2**. (a) and (b) HAADF image of the MnTe/GaAs interface. (c) Fourier filtered image of the interface (b) highlights the misfit regions in yellow. (d) Zoom-in on a region that hosts two misfits. (e) zoom-in on a single misfit region showing a measurement of the spacing between adjacent misfits. (f) High-magnification HAADF image of the interface highlighting the Burgers' circuit in yellow and the Burgers' vector in red. The scale bar is 5nm long.

Fig. 2(a,b) show a high magnification view of the interface taken in HAADF mode, revealing a sharp interface with the substrate. Secondary ion mass spectroscopy measurements were also carried out to characterize the chemistry of the layer across the interface. They are discussed in Appendix A.

A Fourier filtered image of the HAADF measurement is shown at different magnifications in Fig. 2(c-e). It reveals the existence of a misfit array that repeats approximately every 5.5nm to 6nm, releasing lattice strain within two unit-cells from interface. The misfit does not yield any threading dislocations that propagate through the bulk (see Fig. 2(e,f)). Such misfit dislocation arrays are a common feature of heteroepitaxial interfaces and have been observed in GaSb/GaAs [26,27], SnTe/InP [22], PbSe/InAs [28] ZnTe/III-V [29] and CdTe/GaAs interface [30]. They have yet to be observed and studied in epitaxial magnetic materials. Figure 2(b) shows a high-magnification HAADF image of the interface. The yellow arrows point along the $[\bar{1}\bar{1}2]$ direction in GaAs and the $[\bar{1}100]$ direction in MnTe, confirming the epitaxial alignment of the $[11\bar{2}0]_{MnTe}||[1\bar{1}0]_{GaAs}$, identical to the alignment found for MnTe/InP [5,8]. A Burgers' circuit spanning a line that is 6.07nm long (yellow arrows), requires 17 lattice spacings along the $[\bar{1}\bar{1}2]$ direction in GaAs, but 16 spacings along $[\bar{1}100]$ in α-MnTe to complete a full loop. The resulting Burgers' vector is highlighted in red.

**Bulk stacking faults**

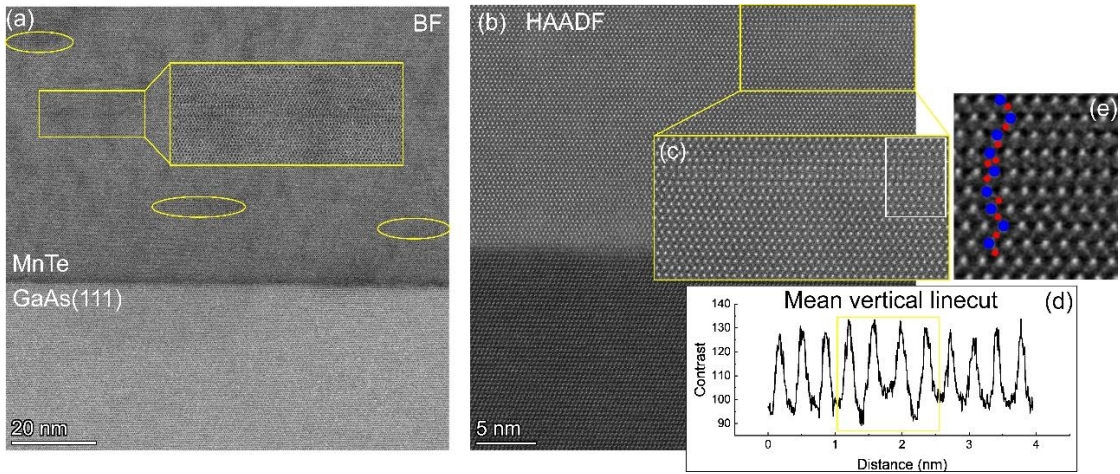

**FIG 3**. (a) Bright field (BF) STEM image of MnTe grown on GaAs highlighting planar bulk defects. The inset is a zoom in on a defective region. (b) HAADF STEM image of the interface at high magnification. (c) Zoom-in on a planar defect. White box is the region where a line-cut is taken. (d) Linecut contrast versus thickness highlighting two anomalous peaks. (e) Atomic structure of the defective region. The defective layer is a misfit where two MnTe crystalline domains of opposite handedness meet.

In Figure 3, we analyze the crystal structure of the layer away from the interface. While we do not observe any threading dislocations, even within frames that are several hundred nm wide (Fig. 1(e)), low magnification BF images such as Fig. 3(a) reveal anomalies in the periodicity of the MnTe structure (see yellow inset). A high magnification image, (Fig. 3(b)), reveals the structure of these anomalous regions. Fig. 3(c) shows a magnified view of such a region, where it is evident that the zigzag periodicity of MnTe is interrupted by a planar stacking fault. We extract a vertical line cut averaged over a region that is 8 atomic columns wide and plot it in Fig. 3(d). It reveals a contrast anomaly as well as a modified peak-to-peak distance (box in Fig. 3(d)) in the c-direction of the MnTe lattice. Fig. 3(e) reveals the fine structure of this defect as a missing Te atomic line, causing a bond between two $Mn_2Te$ dumbbells that do not share a Mn atom. The local chiral structure of the MnTe layer switches handedness on opposite sides of the fault (see Fig. 3(e)). We highlight that the stacking fault that we identify here is not identical to what ref. [8] describes near the interface of MnTe/InP.

It is unclear if and how this defect modifies of the local magnetic structure, a question that is crucial to investigate in future work. The missing Te layer, can be thought of a cluster of Te vacancies. Te vacancies are donors, but the layer studied here was found to be p-type in our previous work as is common for MnTe [5,8,9]. So other acceptor-type defects not visible in TEM must dominate the electric response.

**Surface oxidation**

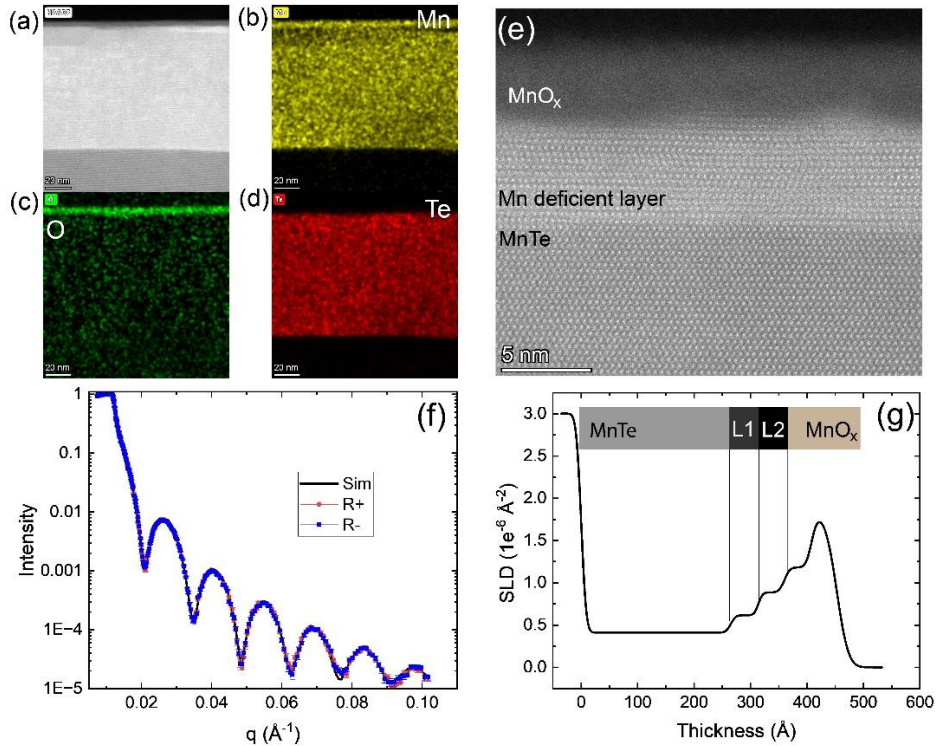

**FIG 4.** (a) TEM image of a region of the 100nm MnTe film where the EDX spectra are acquired. EDX maps of (b) Mn, (c) Oxygen and (d) Te. (e) High magnification TEM image near the sample surface (f) Polarized neutron reflectometry data taken on a 45nm MnTe layer grown on GaAs(111) at 50K, at 1T after field cooling at 1T. Error bars represent data and the solid line is a simulation. (g) Scattering length density used in the simulated reflectometry profile shown in (f). The model excludes magnetic contributions. L1: Defective MnTe layer with a larger SLD than the pristine layer. L2: $MnTe_2$ layer. The $MnO_x$ layer is divided into two layers, with the top layer having a higher density and roughness.

We have also carried out a detailed characterization of the surface of a 100nm layer, that was exposed to atmospheric conditions for several months prior to the measurement. By combining imaging (Fig. 4(a)) with Energy Dispersive X-ray (EDX) mapping in Fig. 4(b-d) we find an amorphous oxide layer approximately 3nm thick on the top of film. The oxide is found to be nearly free of Te (Fig. 4(c,d)) and thus consists in an amorphous $MnO_x$ layer. Underneath the oxide, the MnTe film becomes Mn deficient (Fig. 4(b)). The deficiency results in the formation of isolated $MnTe_2$ layers (bright layer in Fig. 4(a) and bright patches in Fig. 4(e)) beneath the oxide. $MnTe_2$ is a well-studied antiferromagnet [31] that has no net magnetic moment. It is however unknown how two antiferromagets – MnTe and $MnTe_2$ – interact with each other at this interface.

PNR measurements corroborate the oxidation profile seen in TEM. The experiments are performed on the Magnetism Reflectometer [32–34] at the Spallation Neutron Source at Oak Ridge National Laboratory, using neutrons with wavelengths λ in a band of 2.6–8.6 Å and a high polarization of 98.5–99%. Measurements were conducted in a closed cycle refrigerator with a 1T electromagnet. The reflected intensity R+ and R− are measured as a function of wave vector momentum, q = 4πsin(θ)/λ, with the neutron spin parallel (+) or antiparallel (−) to the applied field.

The PNR data shown in Fig. 4(f) is taken at 50K at an applied field of 1T (in-plane) after field cooling at 1T. An uncapped 45nm sample is used for this measurement. A set of five thickness oscillations is observed up to a scattering vector equal to 0.1Å. The data is modeled well using the simulated scattering length density (SLD) profile, shown in Fig. 4(g). The SLD corresponds to the chemical and structural composition of the film [35,36]. The details of the model are shown in Appendix B. The defective surface layer formed as a result of oxidation is found to be close to 24nm thick. Underneath, a pristine 26nm MnTe layer is recovered. The roughness of the top oxide layer is nearly 30% of its thickness in the model. The SLD agrees with the EDX map for Mn (Fig. 4(b)). The PNR spin asymmetry $SA = (R_+ - R_-)/(R_+ + R_-)$ obtained from the experimental reflectivity signals for spin "+" neutrons $R_+$ and spin "-" neutrons $R_-$ is small within error ($\lesssim 0.1$) and is dominated by magnetism originating from the oxide layers. It is discussed in Appendix B. Our findings are consistent with previous findings that a neighboring oxide can alters the composition of the few nearby layers of MnTe in a film [37,38].

**Conclusion**

In summary, we have presented a thorough characterization of the interface, bulk and surface structure of α-MnTe films grown on GaAs(111) by MBE. We have identified a misfit array at the film-substrate interface that plays an important role in releasing strain from the substrate. We have also identified a bulk stacking fault that can potentially alter the local magnetic structure of MnTe. Lastly, we demonstrated that the oxidation of MnTe preferentially impacts Mn atoms, and leaves behind a Te-rich layer that contains $MnTe_2$ islands near the surface beneath the oxide. The interaction of this layer with α-MnTe can influence the magnetic and magnetotransport behavior of ultra-thin films. There are for instance reports of magnetic pinning observed as a shift of the hysteresis loop of MnTe up along the magnetization axis [8]. Such a pinning can be due to interactions between α-MnTe and surface layers that form upon oxidation. Overall, our results highlight the importance of careful control experiments and thorough characterization when isolating signatures of altermagnetism in α-MnTe as well as other candidate magnets. This is particularly relevant for thin films, where the interactions between the substrate, the layer and other unintentional oxide layers can collectively contribute to macroscopic measurements.

**Acknowledgements.** The authors acknowledge support from NSF-DMR-2313441. This research used resources at the Spallation Neutron Source, a Department of Energy Office of Science User Facility operated by the Oak Ridge National Laboratory. SIMS measurements were conducted at the Center for Nanophase Materials Sciences (CNMS), which is a DOE Office of Science User Facility. We also acknowledge support from the Notre Dame Integrated Imaging Facility. This work was supported by Office of Naval Research 6.1 Base Funding at the U.S. Naval Research Laboratory.

**Appendix A. Secondary ion mass spectroscopy**

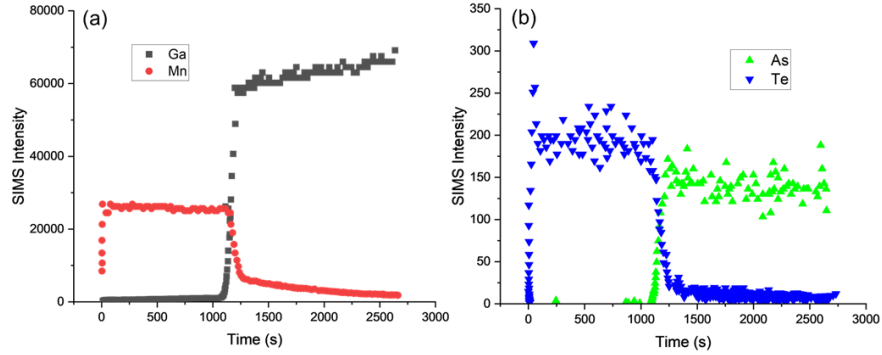

**FIG 5.** Secondary ion mass spectroscopy measurements of the chemical profile of a 150nm MnTe film grown on GaAs. (a) Ga and Mn, (b) As and Te.

**Appendix B. PNR spin-asymmetry and details of the fitting model.**

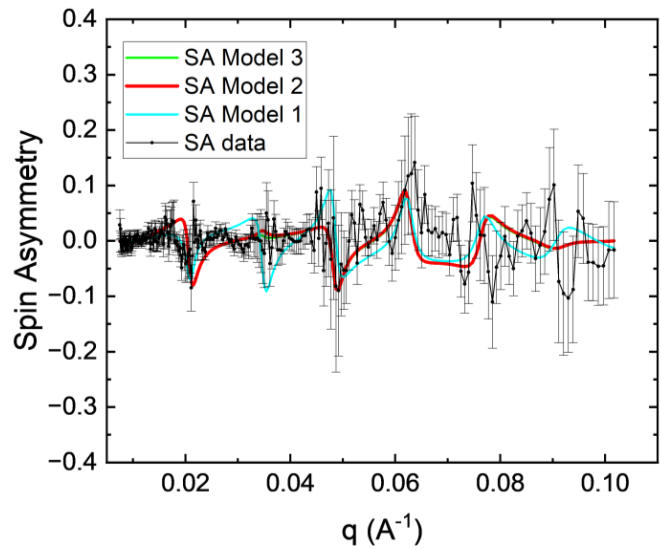

**FIG 6**. Spin asymmetry signal in PNR and the propagated error bar on each data point. The data is acquired at 50K with an in-plane field of 1T. Three models are compared to the data.

| Layer | Thickness (Å) | RMS Roughness (Å) | MSLD model 1 ($10^{-6}$ Å$^{-2}$) | MSLD model 2 ($10^{-6}$ Å$^{-2}$) | MSLD model 3 ($10^{-6}$ Å$^{-2}$) |
|---|---|---|---|---|---|
| GaAs | N/A | 7 | 0 | 0 | 0 |
| MnTe | 265 | 7 | 0 | 0 | 0.003 |
| L1: MnTe | 49 | 6 | 0 | 0.04 | 0.04 |
| L2: MnTe$_2$ | 46 | 8 | 0 | 0.014 | 0.014 |
| MnO$_x$ | 48 | 8 | 0.0434 | 0.031 | 0.031 |
| Roughness MnO$_x$ | 45 | 15 | 0 | 0.008 | 0.008 |

**Table 1.** PNR model parameters shown in Fig. 4(g). The scattering length density is shown in the figure. The PNR data fitting is carried out using GenX version 3.5.9.

The MSLD from three models shown in table 1 is compared to the experimental SA in FIG 7. The SLD is identical for the three models. Model 1 includes magnetism exclusively within the oxide layers, while model 2, includes a finite moment in both the oxide, in the $MnTe_2$ layer and in the defect MnTe layer. Model 3 includes magnetism in all layers. There is an improvement when going from Model 1 to Model 2, however, most of the magnetic signal is contained in the $MnO_x$ and the defective top part of the MnTe layer. Model 3 adds magnetism to the main MnTe layer, but that only improves the fit by a marginal amount. We conclude that most of the magnetic signal detected here stems from defective layers formed after oxidation. It is not possible to isolate the magnetization of the MnTe.

[1]   L. Šmejkal, J. Sinova, and T. Jungwirth, Emerging Research Landscape of Altermagnetism, Phys Rev X **12**, 040501 (2022).

[2]   L. Šmejkal, J. Sinova, and T. Jungwirth, Beyond Conventional Ferromagnetism and Antiferromagnetism: A Phase with Nonrelativistic Spin and Crystal Rotation Symmetry, Phys Rev X **12**, 031042 (2022).

[3]   I. Mazin, Editorial: Altermagnetism—A New Punch Line of Fundamental Magnetism, Phys Rev X **12**, 040002 (2022).

[4]   S.-W. Cheong and F.-T. Huang, Altermagnetism with non-collinear spins, NPJ Quantum Mater **9**, 13 (2024).

[5]   R. D. Gonzalez Betancourt et al., Spontaneous Anomalous Hall Effect Arising from an Unconventional Compensated Magnetic Phase in a Semiconductor, Phys Rev Lett **130**, 036702 (2023).

[6]   K. P. Kluczyk et al., Coexistence of anomalous Hall effect and weak magnetization in a nominally collinear antiferromagnet MnTe, Phys Rev B **110**, 155201 (2024).

[7]   J. D. Wasscher, Evidence of weak ferromagnetism in MnTe from galvanomagnetic measurements, Solid State Commun **3**, 169 (1965).

[8]   M. Chilcote et al., Stoichiometry-Induced Ferromagnetism in Altermagnetic Candidate MnTe, Adv Funct Mater 2405829 (2024).

[9]   S. Bey et al., Unexpected Tuning of the Anomalous Hall Effect in Altermagnetic MnTe thin films, ArXiv 2409.04567 (n.d.).

[10]  J. Ding et al., Large Band Splitting in g-Wave Altermagnet CrSb, Phys Rev Lett **133**, 206401 (2024).

[11]  M. Zeng et al., Observation of Spin Splitting in Room-Temperature Metallic Antiferromagnet CrSb, Advanced Science (2024).

[12]  S. Reimers et al., Direct observation of altermagnetic band splitting in CrSb thin films, Nat Commun **15**, 2116 (2024).

[13]  Y. Guo, H. Liu, O. Janson, I. C. Fulga, J. van den Brink, and J. I. Facio, Spin-split collinear antiferromagnets: A large-scale ab-initio study, Materials Today Physics **32**, 100991 (2023).

[14] I. Mazin, R. González-Hernández, and L. Šmejkal, Induced Monolayer Altermagnetism in MnP(S,Se)$_3$ and FeSe, ArXiv 2309.02355 (2023).

[15] R. B. Regmi et al., Altermagnetism in the layered intercalated transition metal dichalcogenide CoNb4Se8, ArXiv 2408.08835 (2024).

[16] J. Dong, K. Wu, M. Zhu, F. Zheng, X. Li, and J. Zhang, Nonrelativistic spin-splitting multiferroic antiferromagnet and compensated ferrimagnet with zero net magnetization, ArXiv 2501.02914 (n.d.).

[17] R. Watanabe, R. Yoshimi, M. Shirai, T. Tanigaki, M. Kawamura, A. Tsukazaki, K. S. Takahashi, R. Arita, M. Kawasaki, and Y. Tokura, Emergence of interfacial conduction and ferromagnetism in MnTe/InP, Appl Phys Lett **113**, 181602 (2018).

[18] D. Jain, H. T. Yi, A. R. Mazza, K. Kisslinger, M.-G. Han, M. Brahlek, and S. Oh, Buffer-layer-controlled nickeline vs zinc-blende/wurtzite-type MnTe growths on c-plane Al2O3 substrates, Phys Rev Mater **8**, 014203 (2024).

[19] D. Kriegner et al., Magnetic anisotropy in antiferromagnetic hexagonal MnTe, Phys Rev B **96**, 214418 (2017).

[20] G. Kommichau, H. Neumann, W. Schmitz, and B. Schumann, Thermal Expansion of SrF2 at Elevated Temperatures, Crystal Research and Technology **21**, 1583 (1986).

[21] M. Leszczynski, V. B. Pluzhnikov, A. Czopnik, J. Bak-Misiuk, and T. Slupinski, Thermal expansion of GaAs:Te and AlGaAs:Te at low temperatures, J Appl Phys **82**, 4678 (1997).

[22] Q. Zhang, M. Hilse, W. Auker, J. Gray, and S. Law, Growth Conditions and Interfacial Misfit Array in SnTe (111) Films Grown on InP (111)A Substrates by Molecular Beam Epitaxy, ACS Appl Mater Interfaces **16**, 48598 (2024).

[23] S. Vishwanath et al., Comprehensive structural and optical characterization of MBE grown MoSe2 on graphite, CaF2 and graphene, 2d Mater **2**, 24007 (2015).

[24] S. Vishwanath et al., Controllable growth of layered selenide and telluride heterostructures and superlattices using molecular beam epitaxy, J Mater Res **31**, 900 (2016).

[25] R. Minikayev, E. Dynowska, B. Witkowska, A. M. T. Bell, and W. Szuszkiewicz, Unit-cell dimensions of α-MnTe in the 295K - 1200K temperature range, X-Ray Spectrometry **44**, 394 (2015).

[26] M. D. Nordstrom, T. A. Garrett, P. Reddy, J. McElearney, J. R. Rushing, K. D. Vallejo, K. Mukherjee, K. A. Grossklaus, T. E. Vandervelde, and P. J. Simmonds, Direct Integration of GaSb with GaAs(111)A Using Interfacial Misfit Arrays, Cryst Growth Des **23**, 8670 (2023).

[27] S. H. Huang, G. Balakrishnan, A. Khoshakhlagh, A. Jallipalli, L. R. Dawson, and D. L. Huffaker, Strain relief by periodic misfit arrays for low defect density GaSb on GaAs, Appl Phys Lett **88**, 131911 (2006).

[28] B. B. Haidet, E. T. Hughes, and K. Mukherjee, Nucleation control and interface structure of rocksalt PbSe on (001) zincblende III-V surfaces, Phys Rev Mater **4**, 033402 (2020).

[29]  J. Fan, L. Ouyang, X. Liu, D. Ding, J. K. Furdyna, D. J. Smith, and Y. H. Zhang, Growth and material properties of ZnTe on GaAs, InP, InAs and GaSb (0 0 1) substrates for electronic and optoelectronic device applications, J Cryst Growth **323**, 127 (2011).

[30]  F. A. Ponce, G. B. Anderson, and J. M. Ballingall, Interface structure in heteroepitaxial CdTe on GaAs(100), Surf Sci **168**, 564 (1986).

[31]  J. M. Hastings, N. Elliott, and L. M. Corliss, Antiferromagnetic Structures of MnS2, MnSe2, and MnTe2, Physical Review **115**, 13 (1959).

[32]  V. Lauter, H. Ambaye, R. Goyette, W. T. Hal Lee, and A. Parizzi, Highlights from the magnetism reflectometer at the SNS, Physica B Condens Matter **404**, 2543 (2009).

[33]  C. Y. Jiang, X. Tong, D. R. Brown, A. Glavic, H. Ambaye, R. Goyette, M. Hoffmann, A. A. Parizzi, L. Robertson, and V. Lauter, New generation high performance in situ polarized 3He system for time-of-flight beam at spallation sources, Review of Scientific Instruments **88**, 025111 (2017).

[34]  X. Tong et al., In situ polarized 3He system for the Magnetism Reflectometer at the Spallation Neutron Source, Review of Scientific Instruments **83**, 075101 (2012).

[35]  H. J. C. Lauter, V. Lauter, and B. P. Toperverg, *Reflectivity, Off-Specular Scattering, and GI-SAS*, in *Polymer Science: A Comprehensive Reference* (Elsevier, 2012), pp. 411–432.

[36]  V. Lauter-Pasyuk, Neutron Grazing Incidence Techniques for Nano-science., Collect. Soc. Fr. Neutron **7**, s221 (2007).

[37]  S. Siol, Y. Han, J. Mangum, P. Schulz, A. M. Holder, T. R. Klein, M. F. A. M. Van Hest, B. Gorman, and A. Zakutayev, Stabilization of wide band-gap p-type wurtzite MnTe thin films on amorphous substrates, J Mater Chem C Mater **6**, 6297 (2018).

[38]  S. Mori, S. Hatayama, Y. Shuang, D. Ando, and Y. Sutou, Reversible displacive transformation in MnTe polymorphic semiconductor, Nat Commun **11**, 85 (2020).